\documentclass[twocolumn,groupedaddress,showpacs,nofootinbib]{revtex4}
\usepackage{graphicx}
\usepackage{amsmath}
\usepackage{epsfig}

\begin{document}

\title{Dirac Neutrinos, Dark Energy and Baryon Asymmetry}

\author{Pei-Hong Gu$^{1}_{}$}
 \email{pgu@ictp.it}

\author{Hong-Jian He$^{2}_{}$}
\email{hjhe@tsinghua.edu.cn}

\author{Utpal Sarkar$^{3}_{}$}
 \email{utpal@prl.res.in}

\affiliation{$^{1}_{}$The Abdus Salam International Centre for
Theoretical Physics, Strada Costiera 11, 34014 Trieste, Italy\\
$^{2}_{}$Center for High Energy Physics, Tsinghua University,
Beijing 100084, China\\
$^{3}_{}$Physical Research Laboratory, Ahmedabad 380009, India}

\begin{abstract}

We explore a new origin of neutrino dark energy and baryon asymmetry
in the universe. The neutrinos acquire small masses through the
Dirac seesaw mechanism. The pseudo-Nambu-Goldstone boson associated
with neutrino mass-generation provides a candidate for dark energy.
The puzzle of cosmological baryon asymmetry is resolved via
neutrinogenesis.

\end{abstract}


\maketitle

\section{Introduction}

 Strong evidence from cosmological observations \cite{pdg2006}
 indicates that our universe is expanding with an accelerated
 rate at the present. This acceleration can be attributed to
 the dark energy. The dark energy may be a dynamical scalar
 field, such as the quintessence \cite{wetterich1988} with an
 extremely flat potential. The quintessence can be realized
 by a pseudo-Nambu-Goldstone boson (pNGB) arising from
 spontaneous breaking of certain global symmetry near the
 Planck scale \cite{weiss1987}.

On the other hand, various neutrino oscillation experiments
\cite{ms2006} have confirmed that the neutrinos have tiny but
nonzero masses, of the order $10^{-2}\,\textrm{eV}$. The smallness
of neutrino masses can be naturally explained by the seesaw
mechanism \cite{minkowski1977}. In the original seesaw scenario, the
neutrinos are of Majorana nature which, however, has not been
experimentally verified so far. In fact, the ultralight Dirac
neutrinos were discussed many years ago \cite{rw1983,ps1984}.
Recently some interesting models were proposed \cite{mp2002,gh2006},
in which the neutrinos can naturally acquire small Dirac masses,
meanwhile, the observed baryon asymmetry in the universe can be
produced by a new type of leptogenesis \cite{fy1986}, called
neutrinogenesis \cite{dlrw1999}.

It is striking that the scale of dark energy ($\sim\!\!(3\times
10^{-3}\,\textrm{eV})^4$) is far lower than all the known scales in
particle physics except that of the neutrino masses. The intriguing
coincidence between the neutrinos mass scale and the dark energy
scale inspires us to consider them in a unified scenario, as in the
neutrino dark energy model \cite{gwz2003,bgwz2003}. Recently a
number of works studied the possible connection between the pNGB
dark energy and the Majorana neutrinos \cite{bhos2005}.

In this paper, we propose a novel model to unify the mass-generation
of Dirac neutrinos and the origin of dark energy. In particular, a
pNGB associated with the neutrino mass-generation provides the
candidate for dark energy while the neutrino masses depending on the
dark energy field are generated through the Dirac seesaw
\cite{gh2006}. Furthermore, our model also resolves the puzzle of
cosmological baryon asymmetry via the neutrinogenesis
\cite{dlrw1999}.

\section{The Model}

We extend the standard model (SM) gauge symmetry
$SU(2)_{L}^{}\otimes U(1)_{Y}^{}$ with an approximate global
symmetry $U(1)^{3}_{}\equiv U(1)_{1}^{}\otimes U(1)_{2}^{} \otimes
U(1)_{3}^{}$ as well as a discrete symmetry $Z_{2}^{}$. The quantum
number assignment is shown in Table\,\ref{charge}, where $i,j=1,2,3$
denote the family indices, $x_{i}^{}$ is the $U(1)_{i}^{}$ charge,
$\psi_{Li}^{}$ is the left-handed lepton doublet, $\nu_{Ri}^{}$ is
the right-handed neutrino, $H$ and $\eta_{ij}^{}$ are the Higgs
doublets, $\xi_{ij}^{}\equiv \xi_{ji}^{\ast}\,(i\neq j)$ is the
Higgs singlet, $\chi$ is a real scalar. Since all $\eta_{ii}^{}$'s
carry zero $U(1)_{i}^{}$ charge, we only need to introduce one such
doublet-field by defining $\eta_{ii}^{}\equiv \eta_{0}^{}$.\, As for
the other SM fields, which carry even parity under the $Z_{2}^{}$,
they are all singlets under the $U(1)_{i}^{}$ except that the
right-handed charged leptons $\ell_{Ri}^{}$ have the same
$U(1)_{i}^{}$ charge as $\psi_{Li}^{}$. Thus $H$ plays a role of the
SM Higgs.

\begin{table*}
\begin{center}
\begin{tabular}{c|cccc}
\hline\hline
\\[-2.5mm]
Fields          \quad\quad & \quad~~\,$SU(2)_{L}^{}$ \quad\quad & $
U(1)_{Y}^{} \quad\quad $ & $   ~~U(1)_{1}^{}\otimes
U(1)_{2}^{}\otimes U(1)_{3}^{}\quad\quad $& $ Z_{2}^{} $~~\quad\quad
\\[1.5mm]
\hline
\\[-2mm]
$\psi_{Li}^{}$  \quad\quad &     ~~\,\textbf{2}    \quad\quad & $
-1/2 \quad\quad $ & $        x_{i}^{}\times
(\delta_{i1}^{},\delta_{i2}^{},\delta_{i3}) $\quad\quad &$ +
$\quad\quad
\\[1.5mm]
$\nu_{Ri}^{}$   \quad\quad &     ~~\,\textbf{1}    \quad\quad & $
~\,0 \quad\quad $ & $  x_{i}^{}\times
(\delta_{i1}^{},\delta_{i2}^{},\delta_{i3}) $\quad\quad& $  -
$\quad\quad
\\[1.5mm]
$H$         \quad\quad &     ~~\,\textbf{2}    \quad\quad & $ -1/2
\quad\quad $ & $        0        $\quad\quad& $  + $\quad\quad
\\[1.5mm]
$\eta_{ij}^{}$         \quad\quad &     ~~\,\textbf{2} \quad\quad &
$  -1/2 \quad\quad $ & $  x_{i}^{}\times
(\delta_{i1}^{},\delta_{i2}^{},\delta_{i3})-x_{j}^{}\times
(\delta_{j1}^{},\delta_{j2}^{},\delta_{j3})$\quad\quad& $ -
$\quad\quad
\\[1.5mm]
~~~~~~~~~$\xi_{ij}^{}\, (i\neq j)$         \quad\quad &
~~\,\textbf{1} \quad\quad & $ ~\,0 \quad\quad $ & $x_{i}^{}\times
(\delta_{i1}^{},\delta_{i2}^{},\delta_{i3})-x_{j}^{}\times
(\delta_{j1}^{},\delta_{j2}^{},\delta_{j3})$\quad\quad& $ +
$\quad\quad
\\[1.5mm]
$\chi$         \quad\quad &     ~~\,\textbf{1}    \quad\quad & $
~\,0 \quad\quad $ & $  0   $\quad\quad& $  -$\quad\quad
\\[-2.5mm]
\\ \hline \hline
\end{tabular}
\caption{The field content and quantum number assignment. }
\label{charge}
\end{center}
\end{table*}

 The phase transformations of the three Higgs singlets,
 $\xi_{ij}^{}\equiv \xi_{ji}^{\ast}~(i\neq j)$,
 are supposed to be independent and
 hence will result in a global $U(1)^{3}_{}$ symmetry.
 Subsequently, the transformations of
 the Higgs doublets $\eta_{ij}^{}~(i\neq j)$ under this
 $U(1)^{3}_{}$ are determined by requiring the invariance of the
 following scalar interactions,
 \begin{eqnarray}\label{yukawa1}
 \xi_{ij}^{}\chi\eta^{\dagger}_{ij}H+\textrm{h.c.}\,.
 \end{eqnarray}
 However, the six Higgs doublets $\eta_{ij}^{}\,(i\neq j)$
 only have two independent phase transformations to keep
 the following Yukawa interactions invariant,
 \begin{eqnarray}\label{yukawa2}
 \overline{\psi_{Li}^{}}\eta_{ij}^{}\nu_{Rj}^{}+\textrm{h.c.}\,,
 \end{eqnarray}
 which explicitly break the
 $U(1)_{1}^{}\otimes U(1)_{2}^{} \otimes U(1)_{3}^{}$
 down to a $ U(1)_{1}^{'}\otimes U(1)_{2}^{'}$.
 So, in the presence of Eqs.\,(\ref{yukawa1}) and
 (\ref{yukawa2}), we will have two massless Nambu-Goldstone
 bosons (NGBs) and one pNGB after this global symmetry
 is spontaneously broken by the vacuum expectation values
 (\textit{vev}s) of three Higgs singlets $\xi_{ij}^{}$.

We then write down the relevant Lagrangian,
\begin{eqnarray}
\label{lagrangian1} \hspace*{-4mm} -\mathcal{L} &\supset& \!\!\!
\sum_{ij}^{}\left(\rho_{ij}^{2}\right. +\sum_{k\neq
\ell}^{}\lambda_{ij,k\ell}^{}\xi_{k\ell}^{\dagger}\left.\xi_{k\ell}^{}\right)
\eta_{ij}^{\dagger}\eta_{ij}^{}\nonumber\\
&& \!\!\! +\sum_{i\neq j,k\neq \ell,ij\neq k\ell}^{}
\lambda^{'}_{ij,k\ell}\xi_{ij}^{\dagger}
\xi_{k\ell}^{}\eta_{ij}^{\dagger}\eta_{k\ell}^{}
+\left(-\mu_{0}^{}\chi\eta_{0}^{\dagger}
H\right.\nonumber\\
&& \!\!\! +\sum_{i\neq
j}^{}h_{ij}^{}\xi_{ij}^{}\chi\eta_{ij}^{\dagger}H
+\sum_{ij}^{}\left.y_{ij}^{}\overline{\psi_{Li}^{}}\eta_{ij}^{}\nu_{Rj}^{}
+\textrm{h.c.}\right)\!,
\end{eqnarray}
where $\rho_{ij}^{}$ and $\mu_{0}^{}$ have the mass-dimension one
while $\lambda^{(')}_{ij,kl}$, $h_{ij}^{}$ and $y_{ij}^{}$ are
dimensionless. For convenience, we will denote
$\rho_{ii}^{}\equiv\rho_{0}^{}$ and $\lambda_{ii,k\ell}^{}\equiv
\lambda_{0,k\ell}^{}$ corresponding to $\eta_{ii}^{}\equiv
\eta_{0}^{}$.

After the three Higgs singlets $\xi_{ij}^{}$ acquire their
\textit{vev}s, $\langle\xi_{ij}^{}\rangle \equiv
\frac{1}{\sqrt{2}}f_{ij}^{}$, we can write
\begin{eqnarray}
\xi_{ij}^{}= \frac{1}{\sqrt{2}}\left(\sigma_{ij}^{}
+f_{ij}^{}\right)\exp\left(i\varphi_{ij}^{}/f_{ij}^{}\right)\!, &&
(i\neq j),
\end{eqnarray}
with $\sigma_{ij}^{}\,,\,\varphi_{ij}^{}\, (i,j=1,2,3)$ being the
three neutral Higgs and the three NGBs, respectively. Here
$f_{ij}^{}\equiv f_{ji}^{}$, $\sigma_{ij}^{}\equiv \sigma_{ji}^{}$
and $\varphi_{ij}^{}\equiv-\varphi_{ji}^{}$ since $\xi^{}_{ij}\equiv
\xi_{ji}^{\ast}$. In this approach, due to the explicit breaking of
$U(1)_{1}^{}\otimes U(1)_{2}^{} \otimes U(1)_{3}^{}\rightarrow
U(1)_{1}^{'}\otimes U(1)_{2}^{'}$, one of these three NGBs will
acquire a finite mass via the Coleman-Weinberg potential and thus
become a pNGB, while the other two remain massless, as a result of
spontaneous breaking of the subgroup $U(1)_{1}^{'}\otimes
U(1)_{2}^{'}$.

For convenience we redefine the Higgs doublets $\eta_{ij}^{}\,(i
\neq j)$ as
\begin{eqnarray}
\exp\left(i\varphi_{ij}^{}/f_{ij}^{}\right)\eta_{ij}^{}&\rightarrow&\eta_{ij}^{}\,,
\end{eqnarray}
and then express the Lagrangian (\ref{lagrangian1}) in a new form,
\begin{eqnarray}
\label{lagrangian2} -\mathcal{L} &\supset&
M_{0}^{2}\eta_{0}^{\dagger}\eta_{0}^{}+\sum_{i\neq j,k\neq \ell}^{}
\left(M^{2}_{}\right)_{ij,kl}^{}\eta_{ij}^{\dagger}\eta_{k\ell}^{}
\nonumber\\
&& +\left\{-\mu_{0}^{}\chi
\eta_{0}^{\dagger}H+y_{ii}^{}\overline{\psi_{Li}^{}}\eta_{0}^{}\nu_{Ri}^{}\right.
+\sum_{i\neq j}^{}\left[-\mu_{ij}\chi \eta_{ij}^{\dagger}H
\right.\nonumber\\
&&
+\left.\left.y_{ij}^{}\exp\left(-i\varphi_{ij}^{}/f_{ij}^{}\right)
\overline{\psi_{Li}^{}}\eta_{ij}^{}\nu_{Rj}^{}\right]+\textrm{h.c.}\right\}
\end{eqnarray}
with the definitions,
\begin{eqnarray}
\label{massdoublet1} M_{0}^{2}&\equiv&\rho_{0}^{2}+\sum_{k\neq\ell}
\lambda_{0,k\ell}^{}f_{k\ell}^{2}\,,
\\
\label{massdoublet2} (M^{2}_{})_{ij,k\ell}^{}
&\equiv&\left(\rho_{ij}^{2}\right.+\frac{1}{2}\sum_{m\neq
n}^{}\lambda_{ij,mn}^{} \left.f_{mn}^{2}\right)\delta_{ij,k\ell}^{}
\nonumber\\
&& +\frac{1}{2}\lambda^{'}_{ij,k\ell}f_{ij}^{}f_{k\ell}^{}
(1-\delta_{ij,k\ell}^{})\,,
\\
\label{cubic} \mu_{ij}^{}&\equiv &
-\frac{1}{\sqrt{2}}h_{ij}^{}f_{ij}^{}\,,
\end{eqnarray}
At this stage, the last Yukawa term in (\ref{lagrangian2}) depends
on all three fields $\varphi_{ij}$. However, by making the further
phase rotations on the left-handed lepton doublets and the
right-handed neutrinos, we can find that except one combination of
$\varphi_{ij}^{}$ still remains in the Yukawa interaction, the other
two disappear from (\ref{lagrangian2}), so they only have derivative
interactions and stay as the massless NGBs. For instance, we can
make the following rotations,
\begin{eqnarray}
\exp(-i\varphi_{12}^{}/f_{12}^{})\psi_{L2}&\rightarrow&\psi_{L2}^{}\,,\\
\exp(-i\varphi_{12}^{}/f_{12}^{})\nu_{R2}&\rightarrow &\nu_{R2}^{}\,,\\
\exp(+i\varphi_{31}^{}/f_{31}^{})\psi_{L3}&\rightarrow &\psi_{L3}^{}\,,\\
\exp(+i\varphi_{31}^{}/f_{31}^{})\nu_{R3}&\rightarrow
&\nu_{R3}^{}\,,
\end{eqnarray}
and then obtain
\begin{eqnarray}
-\mathcal{L}_{Y}&=&
\sum_{ij}^{}Y_{ij}^{}\overline{\psi_{Li}^{}}
\eta_{ij}^{}\nu_{Ri}^{}+\textrm{h.c.}\,,
\end{eqnarray}
where
\begin{eqnarray}
\label{yukawa3}
Y ~\equiv~ \left\lgroup
\begin{array}{lll}
y_{11}^{} & ~y_{12}^{} & ~y_{13}^{} \\
y_{21}^{}& ~y_{22}^{} & ~y_{23}^{}e^{-i\phi/f}_{}\\
y_{31}^{}& ~y_{32}^{}e^{+i\phi/f}_{} & ~y_{33}^{}
\end{array} \right\rgroup
\end{eqnarray}
with the definition
\begin{eqnarray}
\label{combination} \phi / f &\equiv& \varphi_{12}/f_{12} +
\varphi_{23}/f_{23}  + \varphi_{31}/f_{31}\,.
\end{eqnarray}
Here $f$ should be of the order of the $U(1)_{1}^{'}\otimes
U(1)_{2}^{'}$ breaking scales, i.e., $f\sim f_{ij}^{}$. It is
impossible to remove $\phi$ from the Yukawa interactions by further
transformation. We will show later that this $\phi$ is a pNGB with a
tiny mass and can naturally serve as the candidate of dark energy.

\section{Neutrino Masses}

We consider the case that after the $U(1)^{'}_{1}\otimes
U(1)^{'}_{2}$ breaking, the mass-square (\ref{massdoublet1}) of
$\eta_{0}^{}$ and the eigenvalues of the mass-square matrix
(\ref{massdoublet2}) for $\eta_{ij}^{}\textrm{'s}\,(i\neq j)$ are
all positive. So, these Higgs doublets can develop nonzero
\textit{vev}s only after the SM Higgs-doublet $H$ and the real
scalar $\chi$ both acquire their \textit{vev}s \cite{gh2006},
\begin{eqnarray}
\label{doubletvev} \hspace*{-4mm} \langle\eta_{ij}^{}\rangle \simeq
\left\{
\begin{array}{ll}
\displaystyle \langle H \rangle \langle \chi \rangle\sum_{k\neq
\ell}^{}\left(M^{-2}_{} \right)_{ij,k\ell}^{}\mu^{}_{k\ell}\,,&
~~\textrm{for}~ i\neq j\,,
\\[3mm]
\langle H \rangle \langle \chi \rangle M^{-2}_{0} \mu^{}_{0}\,, &
~~\textrm{for}~ i=j\,.
\end{array}\right.
\end{eqnarray}
In consequence, the neutrinos obtain small Dirac masses,
\begin{eqnarray}
\label{diracmass} \left(m_{\nu}^{}\right)_{ij}^{} & \simeq &
Y_{ij}^{}\langle\eta_{ij}^{}\rangle\,.
\end{eqnarray}

 The discrete $Z_{2}^{}$ symmetry is expected to break
 at the TeV scale by the \textit{vev} of the real scalar
 $\chi$\,\footnote{It
 is also possible to replace the $Z_{2}^{}$ symmetry by a
 global or local $U(1)_X$ symmetry \cite{gh2006,asz2007},
 under which all SM particles transform as singlets,
 while $\nu_{Ri}^{}$, $\eta_{ij}^{}$ and $\chi$ carry the
 $U(1)$ charge $-\frac{1}{2}$, $+\frac{1}{2}$
 and $+\frac{1}{2}$, respectively.
 This $U(1)_X$ symmetry is spontaneously broken at
 TeV by the \textit{vev} of $\chi$.
 So $\langle\chi\rangle$ is fixed by the
 $U(1)_X$ symmetry breaking scale around $O({\rm TeV})$.
 In the case of a local $U(1)_X$ symmetry,
 we need add three massless
 left-handed singlet fermions,
 $s_{Li}^{}\,(i=1,2,3)$ with the $U(1)$ charge $+\frac{1}{2}$,
 which decouple from everything and make the
 theory anomaly free.
 In this case, the new gauge boson couples to
 $s_{Li}^{}$, $\nu_{Ri}^{}$, $\eta_{ij}^{}$ and $\chi$ rather than
 the SM particles,
 so it is expected to escape the detection
 at the LHC and ILC.},
 so we will set $\langle\chi\rangle$ around
 $\mathcal{O}(\textrm{TeV})$\,\footnote{Here we are not
 concerned with the naturalness issue of scalar masses
 as in any non-supersymmetric model.}.
 Furthermore, it is reasonable to take $\mu$ less than $M$ in
(\ref{doubletvev}). Under this setup, it is straightforward to see
that the Dirac neutrino masses will be efficiently suppressed by the
ratio of the electroweak scale over the heavy masses. For instance,
we find that, for $\langle H\rangle\simeq 174\,\textrm{GeV}$,
$M_{}^{}\sim 10^{14}_{} \,\textrm{GeV}$, $\mu \sim 10^{13}_{}
\,\textrm{GeV}$ and $Y \sim \mathcal{O}(1)$, the neutrino masses can
be naturally around $\mathcal{O}(0.1 \,\textrm{eV})$. We see that
this mechanism of the neutrino mass generation has two essential
features: (i) it generates Dirac masses for neutrinos, and (ii) it
retains the essence of the conventional seesaw \cite{minkowski1977}
by making the neutrino masses tiny via the small ratio of the
electroweak scale over the heavy mass scale. This is a realization
of Dirac Seesaw \cite{gh2006}.

 \section{Dark energy}

 So far cosmological observations \cite{pdg2006} strongly
 support the existence of dark energy which accelerates
 the expansion of our universe.
 One plausible explanation for the dark energy has its
 origin in a dynamical scalar field, such as the quintessence
 \cite{wetterich1988} with an extremely flat potential.
 It was shown \cite{weiss1987} that the pNGB provides an
 attractive realization of the quintessence field.

 We have pointed out that after the Higgs singlets getting
 their \textit{vev}s,  one NGB $\phi$
 [as shown in (\ref{yukawa3})-(\ref{combination})] will remain
 in the neutrino Yukawa interactions (which explicitly
 breaks the global $U(1)^3$). Therefore, this NGB will develop
 a finite mass from the Coleman-Weinberg effective potential
 via these neutrino Yukawa interactions,
 and thus become a pNGB.
 We can explicitly compute the Coleman-Weinberg
 potential for $\phi$ at one-loop order,
 \begin{eqnarray}
 \label{V-Q} V(\phi) &=&
 -\frac{1}{16\pi^{2}_{}}\sum_{k=1}^{3}m_{k}^{4}\ln
 \frac{m_{k}^{2}}{\Lambda^{2}_{}}\,,
 \end{eqnarray}
 where $m_{k}^{}$ (as a function of $\phi$) is the \textit{k}th
 eigenvalue of the neutrino mass matrix $m_{\nu}^{}$ and $\Lambda$ is
the ultraviolet cutoff. Note that there is an irrelevant quadratical
term in $V$,
$\frac{\Lambda^{2}_{}}{16\pi^{2}_{}}\sum_{k}^{}m_{k}^{2} $, which
has no $\phi$-dependence and is thus omitted in (\ref{V-Q}). A
typical term in $V$ contributing to the potential of a pNGB field
$Q$ has the form,
\begin{eqnarray}
V(Q) &\simeq& V_{0}^{}\cos (Q/f) \,,
\end{eqnarray}
with $V_{0}^{}=\mathcal{O}(m_{\nu}^{4})$. It is well-known that with
$f$ of the order of Planck scale $M_{\textrm{Pl}}^{}$, $Q$ obtains a
mass of $\mathcal{O}(m_{\nu}^{2}/M_{\textrm{Pl}}^{})$ and is a
consistent candidate for the quintessence dark energy.

\section{Baryon asymmetry}

We now demonstrate how to generate the observed baryon asymmetry in
our model. We make use of the neutrinogenesis mechanism
\cite{dlrw1999}. Since the sphalerons \cite{krs1985} have no direct
effect on the right-handed fields, a nonzero lepton asymmetry stored
in the right-handed fields could survive above the electroweak phase
transition and then produce the baryon asymmetry in the universe,
although the lepton asymmetry stored in the left-handed fields had
been destroyed by the sphalerons. For all the SM species, the Yukawa
couplings are sufficiently strong to rapidly cancel the stored left-
and right-handed lepton asymmetry. But the effective Yukawa
interactions of the Dirac neutrinos are exceedingly weak, and the
equilibrium between the left-handed lepton doublets and the
right-handed neutrinos will not be realized until temperatures fall
well below the electroweak scale. At that time the lepton asymmetry
stored in the left-handed lepton doublets has already been converted
to the baryon asymmetry by the sphalerons. In particular, the final
baryon asymmetry should be
\begin{eqnarray}
B ~=\, \frac{28}{79}\left(B-L_{SM}^{}\right)
          \,=\, \frac{28}{79}L_{\nu_{R}^{}}^{} \,,
\end{eqnarray}
for the SM with three generation fermions and one Higgs doublet.

\begin{figure*} \vspace{2.2cm}
\epsfig{file=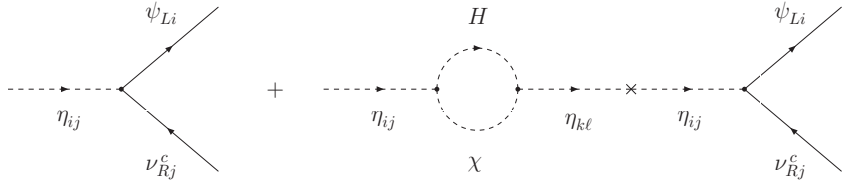, bbllx=6.0cm, bblly=7.0cm, bburx=16cm,
bbury=17cm, width=6.1cm, height=6.2cm, angle=0, clip=0}
\vspace{-5.9cm} \caption{\label{decay} The Higgs doublets decay into
the leptons at one-loop order. Here $i\neq j $, $k\neq \ell$ and
$ij\neq k\ell$. }
\end{figure*}

There are two types of final states coexisting in the decays of
every heavy Higgs doublet,
\begin{eqnarray}
\eta_{ij}^{} \rightarrow
\left\{ \begin{array}{l}
\psi_{Li}^{}\,\nu_{Rj}^{\,c}\,,  \\[2mm]
\chi\,H\,.
\end{array} \right.
\end{eqnarray}
The channels of $\eta \rightarrow \psi_{L}^{}\,\nu_{R}^{\,c}$ and
$\eta^{\ast}_{} \rightarrow \psi_{L}^{\,c}\,\nu_{R}^{}$ can provide
the expected asymmetry between the right-handed neutrinos and
anti-neutrinos if the CP is not conserved and the decays are out of
thermal equilibrium. As shown in Fig.\,\ref{decay}, the mixing
(\ref{massdoublet2}) among $\eta_{ij}^{}\,(i\neq j)$ help to
generate the interference between the tree-level decay and the
irreducible loop-correction. For convenience, we adopt the following
definitions,
\begin{eqnarray}
\widehat{\eta}_{a}^{}&\equiv &\sum_{i\neq j}^{}U_{a,ij}^{}\eta_{ij}^{}\,,
\\
\widehat{M}_{a}^{2}&\equiv &
\sum_{i\neq j,k\neq l}^{}U_{a,ij}^{}\left(M^{2}_{}\right)_{ij,kl}^{}U^{}_{a,kl}\,,
\\
\widehat{\mu}_{a}^{}&\equiv &\sum_{i\neq
j}^{}U_{a,ij}^{}\mu_{ij}^{}\,,
\\
\widehat{Y}_{a,ij}&\equiv & U_{a,ij}^{}Y_{i j}^{}\,,\quad
~~(\textrm{for}\quad i\neq j)\,,
\end{eqnarray}
where $U$ is the orthogonal rotation matrix to diagonalize
$\eta_{ij}^{}\,(i\neq j)$ in their mass-eigenbasis
$\widehat{\eta}_{a}^{}$. We then derive the relevant CP-asymmetry,
\begin{eqnarray}
\label{cpv} \varepsilon_{a}^{}&\equiv&
\frac{\sum_{ij}^{}\left[\Gamma\left(\widehat{\eta}^{\ast}_{a}\rightarrow
\psi_{Li}^{\,c}\nu_{Rj}^{}\right)-\Gamma\left(\widehat{\eta}_{a}^{}\rightarrow
\psi_{Li}^{}\nu_{Rj}^{\,c}\right)\right]}{\Gamma_{a}^{}}\nonumber\\
&=&\frac{1}{4\pi}\sum_{b\neq a}^{}\frac{\textrm{Im}\left[
(\widehat{Y}^{\dagger}_{}\widehat{Y})_{ba}^{}\widehat{\mu}_{b}^{\ast}\widehat{\mu}_{a}^{}\right]}
{(\widehat{Y}^{\dagger}_{}\widehat{Y})_{aa}^{}\widehat{M}_{a}^{2}+\left|\widehat{\mu}_{a}^{}\right|^{2}_{}}\frac{\widehat{M}_{a}^{2}}
{\widehat{M}_{a}^{2}-\widehat{M}_{b}^{2}}\,
\end{eqnarray}
with
\begin{eqnarray} \Gamma_{a}^{}=
\frac{1}{16\pi}\left[(\widehat{Y}^{\dagger}_{}\widehat{Y})_{aa}^{}
+\frac{\left|\widehat{\mu}_{a}^{}\right|^{2}_{}}{\widehat{M}_{a}^{2}}\right]
\widehat{M}_{a}^{}\,
\end{eqnarray}
being the total decay width of \,$\widehat{\xi}_{a}^{}$\, or
\,$\widehat{\xi}_{a}^{\ast}$\,.

For illustration, we will use $\widehat{\xi}_{a}^{}$ to denote the
lightest one among all heavy Higgs doublets (including
$\eta_{0}^{}$), and hence the contribution of $\widehat{\xi}_{a}^{}$
is expected to dominate the final baryon asymmetry, which is given
by the approximate relation \cite{kt1980},
\begin{eqnarray}
\label{asymmetry} Y_{B}^{} \,\equiv\, \frac{n_{B}^{}}{s}\,\simeq\,
\frac{28}{79} \!\times\! \left\{
\begin{array}{lll}
\displaystyle \frac{\varepsilon_{a}^{}}{g_{\ast}^{}}\,, &
~\textrm{for} &K \ll
1,\\[4mm]
\displaystyle \frac{0.3\,\varepsilon_{a}^{}}{g_{\ast}^{}K\left(\ln
K\right)^{0.6}_{}}\,, & ~\textrm{for}&K \gg 1 .
\end{array} \right.
\end{eqnarray}
Here the parameter $K$ is defined as
\begin{eqnarray}
\label{parameter}
K\equiv\left.\frac{\Gamma_{a}^{}}{2H(T)}\right|_{T=\widehat{M}_{a}^{}}
=\left(\frac{45}{16\pi^{3}_{}g_{\ast}^{}}
\right)^{\frac{1}{2}}_{}\frac{M_{\textrm{Pl}}^{}\Gamma_{a}^{}}{\widehat{M}_{a}^{2}}
\end{eqnarray}
which characterizes the deviation from equilibrium. For instance,
inputting $\,\widehat{M}_{a}^{}=
0.1\widehat{M}_{b}^{}=10^{14}_{}\,\textrm{GeV},\,~
|\widehat{\mu}_{a}^{}|=|\widehat{\mu}_{b}^{}|=
10^{13}_{}\,\textrm{GeV},\, ~\left|\sum_{b\neq
a}^{}(\widehat{Y}^{\dagger}_{}\widehat{Y})_{ba}^{}\right|=1.5,
\,(\widehat{Y}^{\dagger}_{}\widehat{Y})_{aa}^{}= 1$,\, and the
maximum CP-phase,  we derive the sample predictions: \,$K\simeq
60$\, and $\,\varepsilon_{a}^{} \simeq 8.0\times 10^{-6}_{}$\,,
where we have used \,$g_{\ast}^{}\sim 100$\, and
$\,M_{\textrm{Pl}}\sim 10^{19}_{}\,\textrm{GeV}$.\, In consequence,
we deduce, $\,n_{B}^{}/s\simeq 10^{-10}$,\, as desired.

\section{Summary and Discussions}

In this paper, we have proposed a new model to realize Dirac
neutrinos, dark energy and baryon asymmetry. In particular, the
heavy Higgs doublets develop small \textit{vev}s which make the
neutrinos acquire small masses through the Dirac seesaw.
Furthermore, the pNGB associated with the Dirac neutrino
mass-generation can be the quintessence field and thus provide an
attractive candidate for dark energy. Finally, our model generates
the matter-antimatter asymmetry in the universe via the
out-of-equilibrium decays of the heavy Higgs doublets with
CP-violation.

In our model, the Dirac neutrino masses are functions of the dark
energy field. The dark energy is a dynamical component and will
evolute with time and/or in space.  In consequence, the Dirac
neutrino masses are variable, rather than constant. The prediction
of the neutrino-mass variation could be verified in the experiments,
such as the observation on the cosmic microwave background and the
large scale structures \cite{bbmt2005}, the measurement of the
extremely high-energy cosmic neutrinos \cite{rs2006} and the
analysis of the neutrino oscillation data \cite{knw2004}.

Finally, we note that the real scalar $\chi$ has a \textit{vev}
around the electroweak scale, it can mix with and couple to the SM
Higgs boson via the quartic interaction,
\begin{eqnarray}
\hspace*{-4mm} \kappa_{\rm eff}^{}\chi^{2}_{}H^{\dagger}_{}H\equiv
\left[\kappa-\left(\sum_{a}^{}\frac{|\widehat{\mu}_{a}^{}|^{2}_{}}{\widehat{M}_{a}^{2}}
+\frac{|\mu_{0}^{}|^{2}_{}}{M_{0}^{2}}\right)\right]\chi^{2}_{}H^{\dagger}_{}H
,
\end{eqnarray}
with $\kappa$ being a dimensionless parameter.  Hence, the SM Higgs
boson is no longer a mass-eigenstate, and its collider signatures
will be modified \cite{bgm1977}. Further phenomenological analyses
for such non-standard Higgs boson will be given elsewhere.

\vspace*{2mm} \textbf{Acknowledgments}: P.H.G. would like to thank
Jun-Bao Wu and Yao Yuan for helpful discussions.

\end{document}